\documentstyle[12pt]{article}
\begin{document}
\def\sqr#1#2{{\vcenter{\hrule height.4pt\hbox{\vrule width.8pt height#2pt 
\kern#1pt\vrule width.8pt}\hrule height.4pt}}} 

\def\Square{\mathchoice{\sqr78\,}{\sqr78\,}\sqr{20.0}{18}\sqr{20.0}{18}}


\def\spose#1{\hbox to 0pt{#1\hss}}\def\lta{\mathrel{\spose{\lower 3pt\hbox
{$\mathchar"218$}}\raise 2.0pt\hbox{$\mathchar"13C$}}}  \def\gta{\mathrel
{\spose{\lower 3pt\hbox{$\mathchar"218$}}\raise 2.0pt\hbox{$\mathchar"13E$}}} 

\newcommand{\dd}{{\rm d}}
\newcommand{\gsim}{\mathrel{%
   \rlap{\raise 0.511ex \hbox{$>$}}{\lower 0.511ex \hbox{$\sim$}}}}
\newcommand{\lsim}{\mathrel{
   \rlap{\raise 0.511ex \hbox{$<$}}{\lower 0.511ex \hbox{$\sim$}}}}



\title{Quasi gravity in branes}

\author{B.Carter,}
\date{DARC/LUTH, Observatoire de Paris, \\ 92195 Meudon, France.
\\ {\it Contrib. to proc. workshop on\\
Analogue Models for General relativity, Rio, Oct., 2000.}}
\maketitle

\begin{abstract}

In contrast with pseudo-gravitational effects that are 
mathematically analogous but physically quite distinct 
from gravity, this presentation deals with a kind of
quasi-gravitational effect that can act in an asymmetrically
moving brane worldsheet in a manner that approximates
(and in a crude analysis might be physically indistinguishable 
from) the effect that would arise from genuine gravitation, of
ordinary Newtonian type in non-relativistic applications,
and of scalar - tensor (Jordan - Brans - Dicke rather than
pure Einstein) type in relativistic applications.

\end{abstract}

\section{Introduction}\label{I}

Most of the talks at this workshop on Analogue Models of Gravity have 
been concerned with what may be termed pseudo-gravitational effects, 
meaning effects whose mathematical description is more or less analogous 
to that of gravity, but whose physical nature is quite distinct. These 
effects typically involve another Lorentz signature metric that coexists 
with the ordinary spacetime metric (whose deviations from flatness are 
interpretable as corresponding to true gravity) but couples to matter in 
an entirely different way, specifying pseudo light cones that typically 
govern the propagation not of real light, nor gravity,  but of some 
quite independent excitation such as sound.

The purpose of the present contribution is to draw attention to something
rather different, what may be described as quasi-gravitational effects, 
meaning phenomena that affect matter locally in approximately the same 
physical manner as true gravity, even though their origin and detailed 
behaviour may be rather different. In the context of non-relativistic 
Newtonian gravitation theory, the most familiar example of such a 
quasi-gravitational effect is the centrifugal field attributable to 
the rotation of the earth that modifies the locally observable Galilean 
acceleration field by contributing a term that is to be added to the 
strictly gravitational contribution due to the Newtonian inverse square
law attraction due to the terrestrial matter distribution. Although 
this centrifugal quasi-gravitational contribution is indistinguishable 
from the truly gravitational contribution in a crude laboratory 
experiment, the difference is of course detectable, via the Coriolis 
effect, in more sensitive experiments such as that of the Foucault 
pendulum.

The kind of quasi gravitational effect I wish to describe here is 
something that modifies the induced spacetime metric on the (q+1) 
dimensional worldsheet of a q-brane in a higher dimensional background 
in a manner that approximates the effect of true (q+1) dimensional 
gravity, even though its origin and precise nature is essentially 
different. It is of particular potential interest in the currently 
fashionable context of models that represent our 4 dimensional universe 
as a 3 brane in a five dimensional background, though in the kind of 
scenarios that are most commonly envisaged the effect considered here 
would be excluded by the usual assumption of symmetry between the two  
opposite sides of the 3-brane, while even if  the symmetry assumption 
were dropped (as has recently been proposed in cases where the q-brane
worldsheet is coupled to a background gauge (q+1) form~\cite{KK,BC,CU})
the quasi gravitational effect could still be overwhelmed by much stronger 
effects of genuinely gravitational origin (just as the terrestrial 
centrifugal effect is overwhelmed by the centrally directed genuinely 
gravitational attraction).

\section{Equation of motion of brane worldsheet}

The effect to be considered here is derivable directly by perturbing the 
general purpose brane worldsheet equation of motion, which is given
~\cite{C95} in terms of the second fundamental tensor
$K_{\mu\nu}^{\ \ \,\rho}$ of the brane worldsheet and of the 
corresponding worldsheet stress energy density tensor 
$\overline T{^{\mu\nu}}$ of the brane by
\begin{equation}\label{001}\overline T{^{\mu\nu}}K_{\mu\nu}^{\ \ \,\rho}=
\perp^{\!\rho}_{\,\mu}\overline f{^\mu}\, ,\end{equation}
where $\overline f{^\mu}$ is the external force density, if any,
acting on the brane, and $\perp^{\!\rho}_{\,\mu}$ is the orthogonal
projection tensor. The complementary (rank p+1) tangential projection
tensor
\begin{equation}\label{002}\gamma^\mu_{ \nu}= g^\mu_{\ \nu}
- \perp^{\!\mu}_{\,\nu} \, ,\end{equation}
i.e. the first fundamental tensor, defines the tangential covariant
differentiation operator 
\begin{equation}\label{003}\overline\nabla_{\!\mu}=\gamma_\mu^{\, \nu}
\nabla_{\!\nu}\, ,\end{equation}
whose action on the first fundamental tensor defines the second 
fundamental tensor according to the specification
\begin{equation}\label{004}K_{\mu\nu}^{\ \ \,\rho}=\gamma^\sigma_{\ \nu}
\overline\nabla_{\!\mu}\gamma^\rho_{\ \sigma}\, ,\end{equation}
which is such as to ensure the Weingarten symmetry condition
\begin{equation}\label{005}K_{\mu\nu}^{\ \ \,\rho}=K_{\nu\mu}^{\ \ \,\rho}
\, ,\end{equation}
as a worldsheet integrability condition, as well as having the
more obvious tangentiality and orthogonality properties
\begin{equation}\label{006}K_{\mu\nu}^{\ \ \,\sigma}\gamma_\sigma^{\ \rho}
=0=\perp^{\!\lambda}_{\,\mu}K_{\lambda\nu}^{\ \ \,\rho}
\, ,\end{equation}
while its trace
\begin{equation}\label{007} K^\rho=K_\mu^{\ \mu\rho}=\overline
\nabla_{\!\nu}\gamma^{\nu\rho}\, ,\end{equation}
inherits the simple worldsheet orthogonality property
\begin{equation}\label{008} \gamma^\rho_{\ \sigma} K^\sigma
=0\, .\end{equation}

\section{Perturbed worldsheet configuration}

The quasi gravity effect to be considered occurs (in its simplest form) 
when the total surface stress energy tensor $\overline T{^{\mu\nu}}$ 
is dominated by an isotropic (Dirac-Nambu-Goto type) contribution 
specified by a large fixed tension $T_{\!_\infty}$ say, together with a
small additional contribution $\tau^{\mu\nu}$ arising from the effect of 
local fields on the brane (representing the observable matter of the 
universe in brane - world scenarios) in the form
\begin{equation}\label{010} \overline T{^{\mu\nu}}=-T_{\!_\infty}
\gamma^{\mu\nu}+\tau^{\mu\nu}\, ,\end{equation}
in the presence of an external force of the commonly occurring
kind (including a Magnus force on a string and a wind force on a sail) 
that is automatically orthogonal to the worldsheet, $\gamma^\mu_{\ \nu}
\overline f^\nu=0$ so that the orthogonal projection on the right
in (\ref{001}) is superfluous. Then if the observable matter contribution 
$\tau^{\mu\nu}$ were absent, the dynamical equation of motion (\ref{001}) 
would reduce to the simple form
\begin{equation}\label{012} T_{\!_\infty} K^\rho= \overline f^\rho
\, .\end{equation}
Starting from an almost uniform (low curvature) reference configuration of
this kind, one can consider an actual configuration that deviates
from this due to the presence of a  matter distribution
$\tau^{\mu\nu}$ confined within a lengthscale that is relatively
small (compared with the reference curvature scale) 
for which the dominant terms in the dynamical equation obtained 
by perturbation of (\ref{001}) can be seen to be given by an expression
of the form
\begin{equation}\label{013} T_{\!_\infty} \delta K^\rho=\tau^{\mu\nu}
K_{\mu\nu}^{\ \ \,\rho}\, ,\end{equation}
in which the perturbation $\delta K^\rho$ of the curvature is given in 
terms of the Dalembertian wave operator $\overline {\Square}=\overline
\nabla{^\nu}\overline\nabla_{\!\nu}$ of the (q+1) dimensional 
worldsheet metric, and of the surface orthogonal vector field $\xi^\mu$ 
specifying the displacement of the worldsheet, by an expression of 
the form
\begin{equation}\label{014} \delta K^\rho\simeq \overline{\Square} 
\,\xi^\rho \, ,\end{equation}
that is obtained from the general curvature perturbation 
formula~\cite{C95} by retaining only the gradient terms of highest order,
which are the ones  that dominate in the localised (short lengthscale)
limit. 

For a brane worldsheet matter distribution that is approximately specified 
with respect to the relevant tangent rest frame unit vector $\bar u^\mu$
($\bar u^\nu \bar u_\nu=-1$) by a stress-energy density tensor of the 
non-relativistic form
\begin{equation} \tau^{\mu\nu}\simeq \bar\rho \,\bar u^\mu \bar u^\nu
\, ,\end{equation}
in terms of a surface mass density $\bar\rho$ whose space section 
volume integal determines the corresponding total mass $M$ say,
the resulting equation takes the form
\begin{equation}\label{016} T_{\!_\infty}\,\overline{\Square}
\,\xi^\rho\simeq \bar\rho\, \bar u^\mu\bar u^\nu K_{\mu\nu}^{\ \ \,\rho}
\, ,\end{equation}
in which, due to the staticity the (hyperbolic) Dalembertian 
operator will reduce to a Laplacian operator (of elliptic type),
so that for a (q-2) spherically symmetric distribution the solution
will be expressible in terms of the radial distance $r$ from the
center by an expression that for $q\geq 3$ will have the power law
form
\begin{equation}\label{017} T_{\!_\infty}\, \xi^\mu= -{M\over{\rm (q-2)}
\Omega^{\rm [q-1]}\,r^{\rm q -2}}\, a^\mu \, ,\end{equation}
in terms of the rest frame orthogonal worldsheet acceleration
vector $a^\mu= \bar u^\nu\overline\nabla_{\!\nu} \bar u^\mu$ given by
\begin{equation}\label{018} a^\rho= \bar u^\mu\bar u^\nu 
K_{\mu\nu}^{\ \ \,\rho} \, .\end{equation}
For the familiar, experimentally accessible, case of an ordinary 
membrane, with q=2, there will be an analogous formula involving
radial dependence of logarithmic rather than power law type.

\subsection{Quasi gravitational metric perturbations}
 
Under the conditions described in the preceeding section, the brane 
worldsheet geometry characterised by the fundamental tensor 
$\gamma_{\mu\nu}$ will be subject to a corresponding perturbation, 
$\bar h_{\mu\nu}=\delta \gamma_{\mu\nu}$ that will be 
given~\cite{C95} in terms of the second fundamental tensor
of the unperturbed reference state by an expression of the form
\begin{equation}\label{020} \bar h_{\mu\nu}= - K_{\mu\nu}^{\ \ \,\rho}
\xi_\rho \, .\end{equation}
This perturbation will have a time component
\begin{equation}\label{021} \bar h_{_{00}}=\bar u^\mu\bar u^\nu\bar 
h_{\mu\nu} \, ,\end{equation}
given by the formula
\begin{equation}\label{022}\bar h_{_{00}}= -2 a^\rho\xi_\rho
\, ,\end{equation}
while the trace $\bar h^\nu_{\ \nu}$ will be given by an expression 
of the analogous form
\begin{equation}\label{023} \bar h^\nu_{\ \nu}= -2 K^\rho\xi_\rho
\, ,\end{equation} 
Evaluating (\ref{022}) explicitly using (\ref{17}), one sees that is
reducible to an expression of the standard (dimensionally 
generalised~\cite{Dvali}) Newtonian form
\begin{equation}\label{024} \bar h_{_{00}}= {2 {\rm G}_{[q+1]} 
M\over r^{\rm q-2}} \, ,\end{equation}
with the relevant generalised Newton constant given by
\begin{equation}\label{025} {\rm G}_{[q+1]}={1\over {(\rm q-2)} 
\Omega^{[\rm q-1]} T_{\!_\infty} }\, a^\rho a_\rho \, .\end{equation}

Although this mechanism  will thus effectively simulate Newtonian type 
gravitational attraction in so far as its effect on non-relativistic
Keppler type orbits is concerned, it leads to a value for the ratio
$\bar h_{_{00}}/\bar h^\nu_{\ \nu}$ that can be seen from (\ref{022})
and (\ref{023}) to be given by (\ref{017}) as
\begin{equation}\label{026}  {\bar h_{_{00}}\over\bar h^\nu_{\ \nu}} = 
{a^\rho a_\rho\over a^\nu K_\nu}\, ,\end{equation}
which will not in general agree with the prediction of Einstein's
theory. Although the ensuing prediction for the relativistic
behaviour (e.g. of light deflection) will thereby deviate from
that of Einstein's purely tensorial theory of gravity, it gives a 
result that will be shown to be matchable by a more general theory of 
the Jordan-Brans-Dicke type to be described in the next section.

\section{Jordan-Brans-Dicke type theories}

The action integral, 
\begin{equation}\label{1} {\cal I}=\int {\cal L}\, \Vert g\Vert^{1/2} \,
{\rm d}^{(q+1)}x\, , \hskip 1 cm {\cal L}={\cal L}_{_{\rm D}}
+{\cal L}_{_{\rm M}} \end{equation}
for a Jordan-Bran-Dicke type scalar tensor theory~\cite{Dicke} in a (q+1)  
dimensional space-time, with metric $g_{\mu\nu}$ in a Dicke type 
conformal gauge (meaning one in which the weak equivalence principle 
is satisfied) is given is given by a Lagrangian density consisting of  
a Dicke type gravitational contribution ${\cal L}_{_{\rm D}}$ involving 
a dilatonic scalar field $\Phi$ as well as the metric, and an ordinary 
matter contribution  ${\cal L}_{_{\rm M}}$ that is independant of $\Phi$, 
with the Dicke contribution given in terms of a coupling constant 
$\omega_{_{\rm D}}$ by an expression of the form
\begin{equation}\label{3} {\cal L}_{_{\rm D}}=
{1\over 2({\rm q-1})\Omega^{[q-1]}}\Big( \Phi R -
{\omega_{_{\rm D}}\over\Phi}g^{\mu\nu}\Phi_{,\mu}\Phi_{,\nu}\Big)
\, , \end{equation}
where $R$ is the Ricci scalar for the metric $g_{\mu\nu}$ and
$\Omega^{[q-1]}$ is the surface area of the unit (q-1) sphere,
which, for an ordinary 4-dimensional spacetime, with space
dimension q=3, will be given by $\Omega^{[2]}=4\pi$.

In terms of the trace of the  material stress energy density tensor,
\begin{equation}\label{5} T_{_{\rm M}}^{\,\mu\nu}= 
2{\partial {\cal L}_{_{\rm M}}\over g_{\mu\nu}}-{\cal L}_{_{\rm M}} 
g^{\mu\nu}\, , \end{equation}
the scalar wave equation for such a theory will be given
in terms of the Dalembertian operator $\Square
=\nabla_{\!\nu}\nabla^{\nu}$ by
\begin{equation}\label{6} \Square\Phi=\alpha_{_{\rm D}}\,
\Omega^{[q-1]}\, T_{_{\rm M}\,\nu}^{\,\nu} \, ,\end{equation}
in terms of a dilatonic coupling constant $\alpha_{_{\rm D}}$
that is given in terms of the original Dicke constant 
$\omega_{_{\rm D}}$ by
\begin{equation}\label{7} { 1\over\alpha_{_{\rm D}}}=
\omega_{_{\rm D}}+{\rm q\over q-1}\, .\end{equation}

To deal with the gravitational equations, it is convenient to express
the dilatonic amplitude $\Phi$ in terms of some fixed value
$\widehat\Phi$ and of a dimensionless scalar field $\phi$ in the form
\begin{equation} \Phi={\rm e}^{-2\phi}\,\widehat\Phi \, \end{equation}
and to change to what is known as an Einstein gauge by  a conformal
transformation $g_{\mu\nu}\mapsto\widehat g_{\mu\nu}$ that
is specified by setting
\begin{equation}\label{10} g_{\mu\nu}={\rm e}^{2\sigma}\,
\widehat g_{\mu\nu}\, ,\end{equation}
where the field $\sigma$ is given in terms of $\phi$ by by the
proportionality relation
\begin{equation}\label{11} 2\phi =({\rm q-1})\sigma\, .\end{equation}
In terms of the Einstein type conformal gauge the action (\ref{1})
will take the form
\begin{equation}\label{12} {\cal I}=\int\widehat {\cal L} \Vert 
\widehat g\Vert^{1/2} \, {\rm d}^{(q+1)}x\, , \hskip 1 cm 
\widehat {\cal L}=\widehat {\cal L}_{_{\rm D}}
+\widehat {\cal L}_{_{\rm L}} \end{equation}
with the matter contribution given by
\begin{equation}\label{14} \widehat {\cal L}_{_{\rm M}}
={\rm e}^{{(\rm q+1)}\sigma}\, {\cal L}_{_{\rm M}}\, ,
\end{equation}
while $\widehat {\cal L}_{_{\rm D}}$ is given as the sum of an
ordinary Einstein-Hilbert type term and a linear scalar field 
contribution in the form
\begin{equation}\label{15} \widehat {\cal L}_{_{\rm D}}=
{\widehat\Phi\over 2({\rm q-1})\Omega^{[q-1]} }\Big(\widehat R -
{4\over\alpha_{_{\rm D}}}\widehat g{^{\mu\nu}}\phi_{,\mu}\phi_{,\nu}\Big)
\, , \end{equation} 
where $\widehat R$ is the Ricci scalar for the Einstein metric 
$\widehat g_{\mu\nu}$ and $\alpha_{_{\rm D}}$ is the constant given by 
(\ref{7}), while the fixed amplitude $\widehat\Phi$ now acts as the 
inverse of the (dimensionally generalised) Newton constant,
which can be identified as
\begin{equation} \widehat {\rm G}_{[\rm q+1]}={1\over\widehat\Phi}
\, .\end{equation}
In this reformulation there will be a matter stress energy density 
contribution given by
\begin{equation}\label{16} \widehat T^{\,\mu}_{_{\rm M}\, \nu}=
{\rm e}^{{(\rm q+1)}\sigma}\, T^{\,\mu}_{_{\rm M}\, \nu}
\end{equation}
whose trace will act as the source for the linear wave equation for 
$\phi$, which will be expressible in the form
\begin{equation}\label{17} \widehat{\Square}\phi=-{1\over 2}
\Omega^{[\rm q-1]}\,\widehat{\rm G}_{[\rm q+1]}\,\alpha_{_{\rm D}}\,
\widehat T^{\,\rho}_{_{\rm M}\, \rho}\, ,\end{equation}
where $\widehat{\Square}$ is the Dalembertian operator for the Einstein 
metric $\widehat g_{\mu\nu}$. The corresponding Einstein type
gravitational equations will be expressible as
\begin{equation}\label{18} \widehat R_{\mu\nu}-{1\over2}\widehat R\, \widehat 
g_{\mu\nu}={2\over\alpha_{_{\rm D}}}\big(2 \phi_{\,\mu}\phi_{\,\nu}-
\widehat g_{\mu\nu}\,\widehat g{^{\rho\sigma}}\phi_{\,\rho}
\,\phi_{\,\sigma}\big)+ {(\rm q-1)}\Omega^{[q-1]}\widehat {\rm G}_{[q+1]}
\widehat T_{_{\rm M}\mu\nu} \, .\end{equation} 

\section{Linearised local scalar tensor field configurations}

Let us now consider the weak field, low source density, limit in which the 
system can be linearised with respect to the dilatonic perturbation field 
$\phi$ and the Einstein metric perturbation field $\widehat h_{\mu\nu}$
defined relative to a flat Minkowski background metric $\eta_{\mu\nu}$
by setting
\begin{equation}\label{21} \widehat g_{\mu\nu}=\eta_{\mu\nu}+
\widehat h_{\mu\nu}\, .\end{equation}
The equation (\ref{17}) for $\phi$ is already linear as it stands,
while the corresponding linearised Einstein equation for $\widehat
h_{\mu\nu}$ is obtainable from (\ref{18}) in the standard form
\begin{equation}\label{23}\Square\,\widehat h_{\mu\nu}
=-2\Omega^{[\rm q-1]}\,\widehat{\rm G}_{[q+1]}\,
\Big( ({\rm q -1})
\widehat T_{_{\rm M}\mu\nu}-\widehat T^{\,\rho}_{_{\rm M}\, \rho}
 \eta_{\mu\nu}\Big)\, .\end{equation}

The gravitational field that is directly measured by the observation
of Keppler type orbits will not be given by this Einstein type metric
$\widehat g_{\mu\nu}$ (to which, due to the involvement of $\phi$
in the relevant stress energy tensor $\widehat T_{_{\rm M}}^{\,\mu\nu}$,
the usual equivalence principle does not apply) but by the original 
Dicke type metric, $g_{\mu\nu}$, which will be expressible analogously 
to (\ref{21}) by
\begin{equation}\label{25}  g_{\mu\nu}=\eta_{\mu\nu}+
h_{\mu\nu}\, ,\end{equation}
with 
\begin{equation}\label{26} h_{\mu\nu} = \widehat h_{\mu\nu}+ 
2\sigma\, \eta_{\mu\nu}
\, ,\end{equation}
to linear order, by (\ref{10}). Usng the relation ({11}),
it can be seen, by combining the wave equations (\ref{17}) 
and (\ref{23}), that the directly observable metric perturbation
$h_{\mu\nu}$ will be given, to linear order, by
\begin{equation}\label{27}\Square\, h_{\mu\nu}=-2\Omega^{[\rm q-1]}
\,\widehat{\rm G}_{[q+1]}\,\Big( ({\rm q -1})  T_{_{\rm M}\mu\nu}
+ \big(1-\Delta_{_{\rm D}} \big)\,  T^{\,\rho}_{_{\rm M}\, \rho}
\eta_{\mu\nu}\Big)\, .\end{equation}
in which the dilatonic deviation constant is given by
\begin{equation}\label{28} \Delta_{_{\rm D}}={1\over ({q-1})\omega_{_{\rm D}}
+{\rm q}}={\alpha_{_{\rm D}}\over{q-1}} \, .\end{equation}

Due to the presence of the deviation constant $\Delta_{_{\rm D}}$,
the coefficient $\widehat {\rm G}_{_{\rm [q+1]}}$ will not be quite
the same as the effective Newtonian coupling constant 
${\rm G}_{_{[q+1]}}$ that will be observed in the static non-relativistic
limit for which, in terms of the relevant rest frame unit vector
$u^\mu$ (with $u^\mu u_\mu=-1$) the stress-energy density will be 
approximately of the form
\begin{equation}\label{32} T_{_{\rm M}}^{\,\mu\nu}=\rho u^\mu u^\nu
\, , \end{equation}
in which $\rho$ is the mass density, whose space volume integral
will be identifiable in this limit as the total mass $M$. It can be
seen that for a spherically summetric distribution the time component, 
\begin{equation}\label{33} h_{_{00}}= u^\mu u^\nu h_{\mu\nu}
\, ,\end{equation}
of the metric perturbation will be given in terms of the radial
distance $r$ from the center by an expression of the standard
(dimensionally generalised~\cite{Dvali}) Newtonian form
\begin{equation}\label{44} h_{_{00}}= {2 {\rm G}_{_{[q+1]}}M
\over r^{\rm q-2}} \, ,\end{equation}
but with the effective gravitational coupling constant given by
\begin{equation}\label{35} {\rm G}_{_{\rm [q+1]}}=
\widehat {\rm G}_{_{\rm [q+1]}} \Big(1+{\Delta_{_{\rm D}}
\over{\rm q-2}}\Big)\, .\end{equation}
It can be seen that it will be related to the corresponding expression 
for the trace $h^\rho_{\ \rho}$ of the metric perturbation by 
\begin{equation}\label{36} h^\rho_{\ \rho}={2-({\rm q+1})\Delta_{_{\rm D}}
\over {\rm q-2}+\Delta_{_{\rm D}}} \, h_{_{00}} \, ,\end{equation}
which is equivalent to
\begin{equation}\label{37} h_{_{00}}={({\rm q-2})\omega_{_{\rm D}}+{\rm q-1}
\over 2\omega_{{\rm D}}+1}\, h^\rho_{\ \rho}\, .\end{equation}

\section{Conclusion}

It can be seen that the ratio (\ref{37}) can be matched by the
simulation effect leading to the corresponding ratio (\ref{026})
if the relevant reference frame curvature vector $K^\mu$ and
the corresponding acceleration vector $a^\mu$ are related by
\begin{equation}\label{40}
 a^\rho a_\rho={({\rm q-2})\omega_{_{\rm D}}+{\rm q-1}
\over 2\omega_{{\rm D}}+1}\, K^\rho a_{\rho}\, ,\end{equation}
or equivalently
\begin{equation}
a^\rho K_\rho= {2+({\rm q+1})\Delta_{_{\rm D}}\over
{\rm q-2}+ \Delta_{_{\rm D}}}\, a^\rho K_\rho\, .\end{equation}

This means that, in the linear approximation we have been using,
the quasi gravitational effect arising from the extrinsic curvature 
of the brane simulates what would be predicted by a 
Jordan-Brans-Dicke theory with $\omega_{_{\rm D}}$ given by what
is obtained by solving (\ref{40}), namely
\begin{equation}\label{42} \omega_{_{\rm D}}={{(\rm q-1)} a^\rho K_\rho
-a^\rho a_\rho \over  2 a^\nu a_\nu -{(\rm q-2)} a^\nu K_\nu}
\, ,\end{equation}
which corresponds to a dilatonic deviation $\Delta_{_{\rm D}}$
given by
\begin{equation} \Delta_{_{\rm D}}={2 a^\rho a_\rho +{(2-q)} 
a^\rho K_\rho\over {(\rm q+1)} a^\nu a_\nu+ a^\nu K_\nu}
\, .\end{equation}

It is to be emphasised that the approximation presented here
has been derived only for static configurations in a linearised 
weak field limit, and cannot be expected to remain accurate when 
stronger fields or significant deviations from staticity are 
involved.

The author wishes to express thanks to M\'ario Novello for initiating the
organisation of this workshop, to Matt Visser for ensuring the completion 
of the edited proceedings, and to Gilles Esposito-Far\`ese for technical 
discussions about the simulated worldsheet gravity mechanism in the 
context of scalar - tensor theories.

\end{document}